\documentclass[twocolumn,superscriptaddress,showpacs,amsmath,amssymb,pra,longbibliography]{revtex4-1}

\usepackage[pdftex]{graphicx} 
\usepackage{dcolumn}  
\usepackage{bm}       
\usepackage[usenames,dvipsnames]{color}
\definecolor{URLCOL}{rgb}{0,0.52,0.83} 
\definecolor{LINKCOL}{rgb}{0.05,0.5,0} 
\definecolor{orange}{rgb}{0.6,0.3,0} 
\definecolor{CITECOL}{rgb}{0.25,0,0.48} 
\usepackage{epstopdf}

\usepackage[pdftex,bookmarks,breaklinks,bookmarksopen,bookmarksnumbered,colorlinks,linkcolor=LINKCOL,linktocpage,citecolor=CITECOL,urlcolor=URLCOL,pdfpagemode=UseOutline,pdftex]{hyperref}

\usepackage{booktabs}
\usepackage{natbib}

\definecolor{TITLECOL}{rgb}{0.1,0.2,0.7} 
\definecolor{SECOL}{rgb}{0.1,0.2,0.7} 
\definecolor{CONTENTSCOL}{rgb}{0.1,0.2,0.7} 
\definecolor{SSECOL}{rgb}{0.25,0,0.48} 
\definecolor{SSSECOL}{rgb}{0.2,0.08,0.53} 
\definecolor{FINCOL}{rgb}{0.01,0.3,0.07} 
\definecolor{ENSEMBLECOL}{rgb}{0.5,0.1,0.8} 

\def\gs{^\text{gs}}
\def\prltitle#1{{\em #1.--}}
\def\lemma{\prltitle{Lemma}}
\def\theorem{\prltitle{Theorem}}

\def\proof{{\em Proof.--}}

\definecolor{URLCOL}{rgb}{0,0.17,0.43} 
\definecolor{LINKCOL}{rgb}{0.05,0.4,0} 
\definecolor{CITECOL}{rgb}{0.35,0,0.48} 

\def\sss{\scriptscriptstyle\rm}

\def\bea{\begin{eqnarray}}
\def\eea{\end{eqnarray}}
\def\ben{\begin{equation}}
\def\een{\end{equation}}
\def\benu{\begin{enumerate}}
\def\enu{\end{enumerate}}

\def\bei{\begin{itemize}}
\def\eei{\end{itemize}}
\def\beit{\begin{itemize}}
\def\eit{\end{itemize}}
\def\benu{\begin{enumerate}}
\def\enu{\end{enumerate}}

\def\hatT{{\hat T}}
\def\hatV{{\hat V}}

\def\br{{\bf r}}

\def\half{\frac{1}{2}}

\def\s{_{\sss S}}

\def\Hxc{_{\sss HXC}}

\def\ee{_{\rm ee}}

\def\intr{\int d^3r\,}

\def\n{n}

\def\Eqsref#1{Eqs.~\eqref{#1}}
\def\Eqref#1{Eq.~\eqref{#1}}

\def\Figref#1{Fig.~\ref{#1}}
\def\Ref#1{Ref.~\cite{#1}}

\begin{document}

\title{
Guaranteed convergence of the Kohn--Sham equations
}
\author{Lucas O.\ Wagner}
\affiliation{Department of Physics and Astronomy, University of California, Irvine, CA 92697}
\affiliation{Department of Chemistry, University of California, Irvine, CA 92697}
\author{E.M. Stoudenmire}
\affiliation{Department of Physics and Astronomy, University of California, Irvine, CA 92697}
\author{Kieron Burke}
\affiliation{Department of Physics and Astronomy, University of California, Irvine, CA 92697}
\affiliation{Department of Chemistry, University of California, Irvine, CA 92697}
\author{Steven R.\ White}
\affiliation{Department of Physics and Astronomy, University of California, Irvine, CA 92697}
\date{\today}

\begin{abstract}
A sufficiently damped iteration 
of the Kohn--Sham equations with the exact functional is proven to always 
converge to the true ground-state density, regardless of
the initial density or the strength of electron correlation, for finite Coulomb systems.
We numerically implement the exact functional for one-dimensional continuum systems
and demonstrate convergence of the damped KS algorithm.
More strongly correlated systems converge more slowly.
\end{abstract}

\pacs{%
31.15.E-, 
71.15.Mb, 
05.10.Cc 
}

\maketitle



Kohn--Sham density functional theory (KS-DFT) \cite{KS65} is a widely applied 
electronic structure method.
Standard approximate functionals 
yield accurate ground-state energies and electron densities for many systems
of interest \cite{B12}, but often fail
when electrons are strongly correlated.
Ground-state properties can be qualitatively incorrect  \cite{MCY09},
and convergence can be very slow \cite{DS00,TOYJ04}.
To remedy this, several popular schemes augment Kohn--Sham theory, such as LDA+U \cite{AAL97}.
Others seek to improve approximate functionals \cite{HSE06} within the original formulation.
But what if the exact functional does not exist for strongly correlated systems?
Even if it does, what if the method fails to converge?
Either plight would render KS-DFT useless for 
strongly correlated systems, and render fruitless the vast efforts currently
underway to treat e.g., oxide materials \cite{ABLH13}, with KS-DFT.

The Kohn--Sham (KS) approach employs
a fictitious system of non-interacting electrons,
defined to have the same density as the interacting system of interest.
The potential characterizing this KS system 
is unique if it exists \cite{HK64}.
Because the KS potential is a functional of the density, in practice
one must search for the density and KS potential together
using an iterative, self-consistent scheme \cite{DG90}.
The converged density is in principle the ground-state density of the original,
interacting system, whose ground-state energy is 
a functional of this density.

Motivated by concerns of convergence and existence, 
we have been performing KS calculations with the exact functional
for one-dimensional (1d) continuum systems \cite{SWWB12,WSBW12}.
Even when correlations are strong, we never find a density whose KS potential
does not exist, consistent with the results of \Ref{CCR85}.
Nor do we find any system where the KS scheme does not converge,
although convergence can slow by orders of magnitude as correlation is increased,
just as in approximate calculations \cite{DS00,TOYJ04}.

\begin{figure}
\includegraphics[width=\columnwidth]{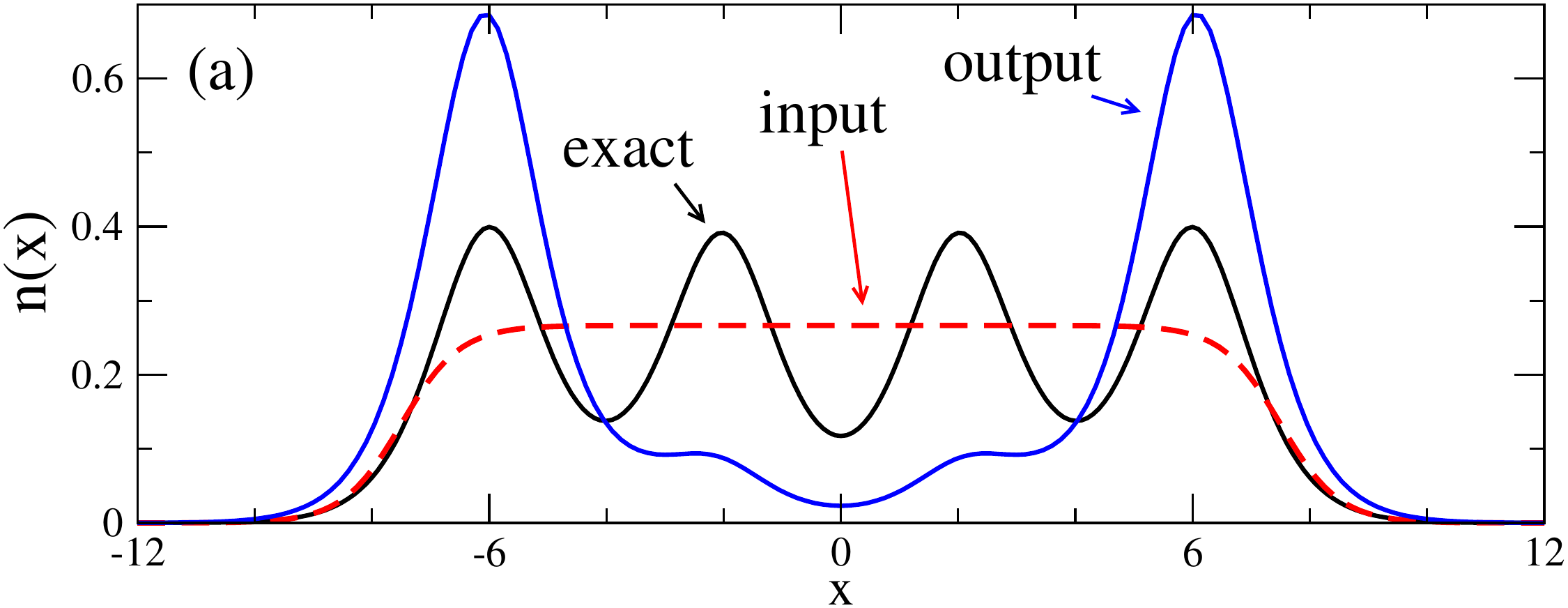}
\includegraphics[width=\columnwidth]{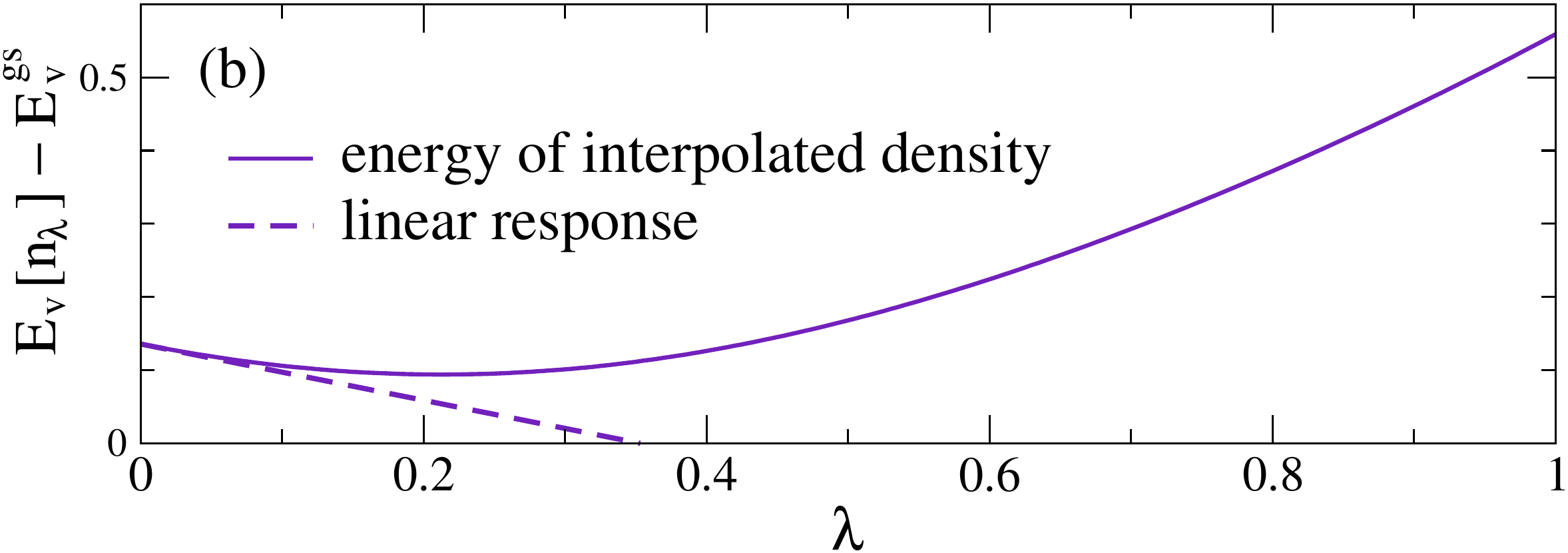}
\caption{
(a) The input and output densities for a single step of the Kohn--Sham
scheme, as well as the exact density, of
 a one-dimensional, strongly correlated 4 atom, 4 electron system.
(b) The energy of the system  which interpolates between the input and output densities, 
$E_v[n_\lambda]$, measured from the ground-state energy $E\gs_v$.
Also shown is the linear-response approximation with slope given by \Eqref{eqn:linearstep}.
 }
\vskip -0.7cm
\label{H4b4inout}
\end{figure}

Exact statements about the unknown density functional inform the
construction of all successful DFT approximations \cite{LP85,PBE96,PK03,PRTS05}.  More importantly,
they distinguish
between what a KS-DFT calculation can {\em possibly} do, and what it cannot.
The most notorious example is the demonstration that the 
KS band-gap of a semiconductor does not equal the true charge gap, even when
the exact functional is used \cite{PPLB82,SWWB12}.  
Our key result 
is an analytic proof that  a simple algorithm guarantees convergence of the KS
equations for all systems, weakly or strongly correlated, independent
of the starting point.  Thus
multiple
stationary points and failures to converge  are artifacts of approximate functionals.
Studies of convergence are well-known in applied mathematics;
but almost all concern simple approximations, such as LDA \cite{AC09}, Hartree-Fock \cite{CL00},
etc., and not those in current use in many calculations.

The basic idea lies in a single
step of the KS scheme, which proceeds from an input density
to produce an output density. For a strongly correlated system as in \Figref{H4b4inout}.a,
the output density can differ strongly from
the input density, and be further from the true ground-state density.
Nevertheless, by proving that the initial slope is always negative as in \Figref{H4b4inout}.b,
we show there is always a linear combination of the input and output densities
that lowers the energy.
By sufficiently damping each KS step, the energy is always reduced each iteration,
yielding the ground-state
density and energy to within a given tolerance in a finite number of iterations.

The KS algorithm is designed to minimize the energy as
a functional of the electron density, $n(\br)$.
For an $N$-electron system with a reasonable \footnote{%
The precise restrictions on potentials are detailed in \Ref{L83};
Coulomb potentials are allowed.}
external potential $v(\br)$, the energy functional is \cite{KS65}:
\ben
E_v[n] = T\s[n] + \intr n(\br)\, v(\br) + E\Hxc[n],\label{KSE}
\een
where 
$T\s[n]$ is the kinetic energy of non-interacting (NI) electrons having density $n(\br)$, 
and $E\Hxc[n]$ is the Hartree-exchange-correlation (HXC) energy \cite{L83,FNM03}.
The KS equations are, in atomic units,
\ben
-\half \nabla^2 \phi_j(\br) + \Big( v(\br) + v\Hxc[n](\br) \Big) \phi_j(\br) = \epsilon_j\, \phi_j(\br), \label{KSeqn}
\een
where $v\Hxc[n](\br) = \delta E\Hxc[n]/\delta n(\br)$ is the HXC potential,
$\phi_j(\br)$ are the electron orbitals, and $\epsilon_j$ their eigenvalues.
(In this work, we consider spin-unpolarized
systems for simplicity.)
An output density $n'(\br)$ is
found by doubly occupying the lowest-energy orbitals:
\ben
n'(\br) = 2\sum_{j=1}^{\infty} f_j \,|\phi_j(\br)|^2, \label{outdens}
\een
where $0 \le f_j \le 1$ and $\sum_j f_j = N/2$.  Fractional occupation is only allowed
for the highest occupied orbitals if they are degenerate, where $f_j$ is chosen 
to minimize the difference between $n(\br)$
and $n'(\br)$ \footnote{\label{orbrot}%
An orbital rotation among degenerate orbitals may also be required;  see \Ref{UK01}.
}. \nocite{UK01}


Consider convergence of the following simple algorithm.
Given an input density $n(\br)$,
solve the KS equations to obtain the output density $n'(\br)$.
Define
\ben
\eta \equiv  \dfrac{1}{N^2} \intr \big( n'(\br) - n(\br) \big)^2.  \label{Dn} 
\een
Choose some small $\delta > 0$, and if $\eta < \delta$, then the
calculation has converged.
Otherwise, the next input is
\ben
n_\lambda(\br) = (1-\lambda)\, n(\br) + \lambda\, n'(\br), \label{lambdadens}
\een
for some $\lambda \in (0, 1]$, and repeat.
An ensemble-$v$-representable $\n(\br)$ is the ground-state density (or an
ensemble mixture of degenerate ground-state densities) for some local
potential $v[\n](\br)$ \cite{L82,Leeuwen}.  
For NI electrons, this potential is $v\s[\n](\br)$.  We
call $\n(\br)$ physical when both potentials exist, and we require
all $n_\lambda(\br)$ to be physical.
We refer to a single iteration of \Eqsref{KSeqn}-\eqref{lambdadens} as one step of the KS algorithm.
Taking full steps with $\lambda=1$ does not usually
lead to a fixed point. But taking
damped steps with $\lambda < 1$ ensures the algorithm converges,
as we now prove.


\lemma
Consider two finite \footnote{%
We restrict our attention to finite systems, since
in an extended system \Eqref{ineq} would be ill defined.} 
systems of $N$ electrons,
with ground-state densities $n(\br)$, $n'(\br)$,
and potentials $v[n](\br)\neq v[n'](\br)$, by which we mean
the potentials differ by more than a constant.
Then \cite{HK64}
\ben
 \intr \Big( v[n^\prime](\br) - v[n](\br) \Big)\big( n^\prime(\br) - n(\br) \big) < 0. \label{ineq}
\een

\proof
Following  \Ref{HK64}, we apply the variational principle.
  Since $n(\br)$ is the
ground-state density of the potential $v[n](\br)$,
we have  $E_{v[n]}[n] < E_{v[n]}[n^\prime] $, or 
\ben
\intr v[n](\br)\big( n(\br) - n^\prime(\br) \big) 
<
F[n^\prime] - F[n], \label{ineq0}
\een
where $F[n] \equiv T\s[n] + E\Hxc[n]$.
It is also true that $E_{v[n^\prime]}[n^\prime] < E_{v[n^\prime]}[n]$,
so we may switch primes with unprimes in \Eqref{ineq0}.  Adding the resulting
equation to the original yields  \Eqref{ineq}. $\square$
Note that the lemma is true for any interaction between electrons,
including none.

\theorem
Given an arbitrary physical  density $n(\br)$ as input into the KS algorithm,
\ben
E_v'[n] \equiv \left. \dfrac{d E_v[n_\lambda]}{d \lambda}\right|_{\lambda = 0} \le 0, \label{thm}
\een
where $n_\lambda(\br)$ is defined as in \Eqref{lambdadens}.
If equality holds, then $n(\br)$ is
a stationary point of $E_v[n]$.

\proof 
Consider $\Delta E_v$  
resulting from 
$\lambda\, \Delta n(\br)$ \mbox{$\equiv \lambda\,  \big(n^\prime(\br) - n(\br)\big)$}
\mbox{$= n_\lambda(\br) - n(\br)$}.
Then
\ben
E_v'[n] = 
\intr \left.\dfrac{\delta E_v[n]}{\delta n(\br)}\right. \Delta n(\br). \label{DEv}
\een
For a physical density,
the functional derivative is \cite{Leeuwen}
\bea
\dfrac{\delta E_v[n]}{\delta n(\br)}
&=&  -v\s[n](\br) + v(\br) + v\Hxc[n](\br). \label{dEn}
\eea
Since \mbox{$ v(\br) + v\Hxc[n](\br)$} defines $v\s[n'](\br)$ 
($n'(\br)$ is the output density of \Eqref{KSeqn}), we have:
\ben
\left.\dfrac{\delta E_v[n]}{\delta n(\br)}\right.
=  v\s[n'](\br) - v\s[n](\br). \label{dEn0}
\een
Combining \Eqref{dEn0} and \Eqref{DEv} gives:
\ben
E_v'[n] = \intr \Big(v\s[n'](\br) -v\s[n](\br) \Big)\,\big(n^\prime(\br) - n(\br)\big).
\label{eqn:linearstep}
\een
Two cases arise:  if $v\s[n'](\br) \neq v\s[n](\br)$, use the lemma applied to
NI systems: then $E_v'[n]$ must be less than zero.
Otherwise, $v\s[n'](\br) = v\s[n](\br)$, so both $E_v'[n]$ 
and the RHS of \Eqref{dEn0} are zero, and
$n(\br)$ is a stationary point of $E_v[n]$. $\square$
We illustrate the theorem in \Figref{H4b4inout}.b, where we plot $E_v[n_\lambda]$ and
its linear-response approximation for the input density
of \Figref{H4b4inout}.a.


\prltitle{Corollary 1}
The KS algorithm described above is guaranteed to converge to a stationary point
 of the functional, if (1) only physical densities are encountered,
(2) the  energy functional is convex, and (3)  appropriate values
for $\lambda$ are used, e.g.\ from the algorithm of \Ref{BMR03},
because it is effectively a gradient-descent algorithm
\footnote{%
With respect to gradient descent minimization, 
we assume that functionals behave like their less dimensionful
counterparts, functions.  A more comprehensive analysis is beyond the
scope of this work.
}. 

\prltitle{Corollary 2}
When using the exact functional, the KS algorithm using appropriate $\lambda$'s converges
to the exact ground-state density, as long as the first input density is a physical density.
This is because we can choose each subsequent input density
as a physical density 
\footnote{
In the language of DFT, we can find an ensemble-$v$-representable density which
is arbitarily close to a density $\bar{n}(\br)$
which has reasonable properties:  integrating to a finite particle number $N$, with a finite von Weizs{\"a}cker kinetic
energy, i.e.\ $\sqrt{\bar{n}(\br)} \in H^1(\mathbb{R}^3)$, and being nowhere negative \cite{L83,Leeuwen}.  
This set of reasonable densities is convex \cite{Leeuwen}; so that $n_\lambda(\br)$ is reasonable if both $n(\br)$ and
$n'(\br)$ are, which is the case.  The examples of \Ref{EE83} are not in this set.
},   \nocite{EE83}
and the exact ensemble-functional \cite{V80,L83} is convex.
The only stationary point of the exact functional,
when considering physical densities, 
is the ground-state density \cite{PL85}.

\prltitle{Numerical implementation}%
To find the KS energy functional exactly when there is no degeneracy, we must find
the many-electron wavefunction $\Psi[n]$ that minimizes $\langle \Psi| \hatT + \hatV\ee | \Psi \rangle$
(the kinetic and electron-electron repulsion energies)
with density $n(\br)$ \cite{L79,L83}.  
To perform this very demanding \cite{SV09} interacting inversion, start with a guess for the potential,
 $\tilde v(\br)$.
Then solve the many-body system for the ground-state wavefunction 
$\tilde \Psi$ and density $\tilde n(\br)$.
Using a quasi-Newton method \cite{B65},  modify
$\tilde v(\br)$ and repeat, minimizing the difference between
$\tilde n(\br)$ and the target density $n(\br)$.
Once converged, the procedure is repeated for NI electrons.
The HXC energy is then
\ben
E\Hxc[n] = \langle \Psi[n] | \hatT + \hatV\ee| \Psi[n] \rangle - T\s[n], \label{EHxc}
\een
and the HXC potential is
\ben
v\Hxc[n](\br) = v\s[n](\br) - v[n](\br).\label{vHxc}
\een
We implement these functionals for 1d continuum systems \cite{SWWB12,WSBW12},
obtaining highly-accurate many-body solutions with the density matrix renormalization group \cite{White:1992,White:1993a}.
These are the first such inversions for systems with more than 2 electrons \cite{TGK08,CCD09}.
Because, in 1d, degeneracy (beyond spin) does not occur, we find pure states $\Psi[n]$.
More generally, one should invert using an ensemble $\Gamma[n]$ and take a trace in \Eqref{EHxc} \cite{V80,L83}.

\begin{figure}
\includegraphics[width=\columnwidth]{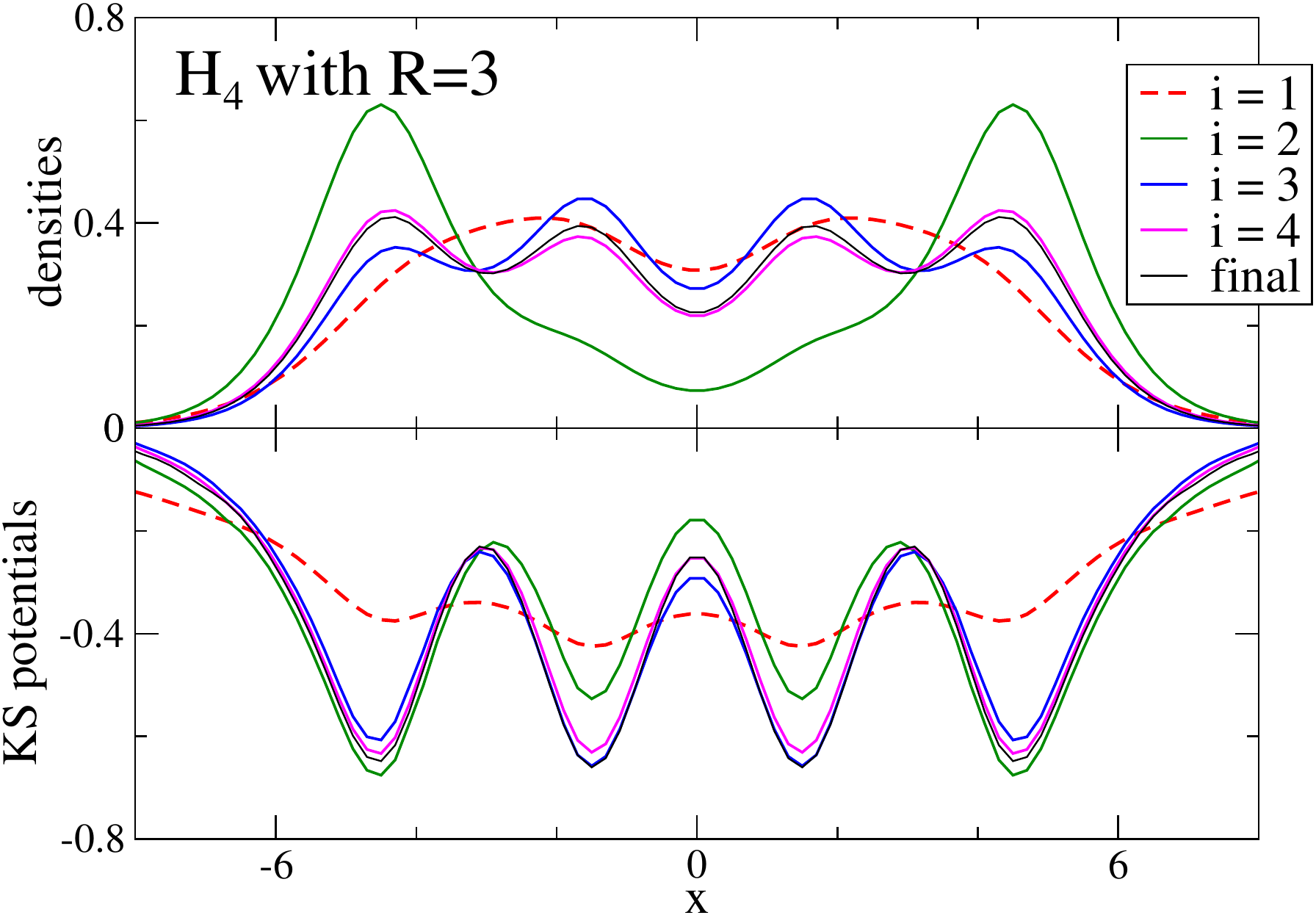}
\caption{KS procedure for a moderately correlated 4-electron system
(four hydrogen atoms with $R=3$),
showing the first few iterations.  Using a fixed $\lambda = 0.30$, 
we converge to $\eta < 10^{-6}$ using \Eqref{Dn} within 13 iterations.
 }
\label{H4b3KS}
\vskip -0.5cm
\end{figure}

To illustrate convergence of the damped KS algorithm
using the exact functional, we plot the output densities and KS potentials
for a four-electron, four-atom system in \Figref{H4b3KS}.  
We choose the interatomic spacing $R=3$ to be roughly twice the equilbrium spacing of H$_2$
(when the interaction between nuclei is the same as that between electrons),
making this a moderately correlated system.
Taking  $\lambda = 0.30$, the algorithm converges to the exact density
(computed separately using DMRG)
to $\eta < 10^{-6}$ using \Eqref{Dn}, within 13 steps.

\begin{figure}[h]
\includegraphics[width=\columnwidth]{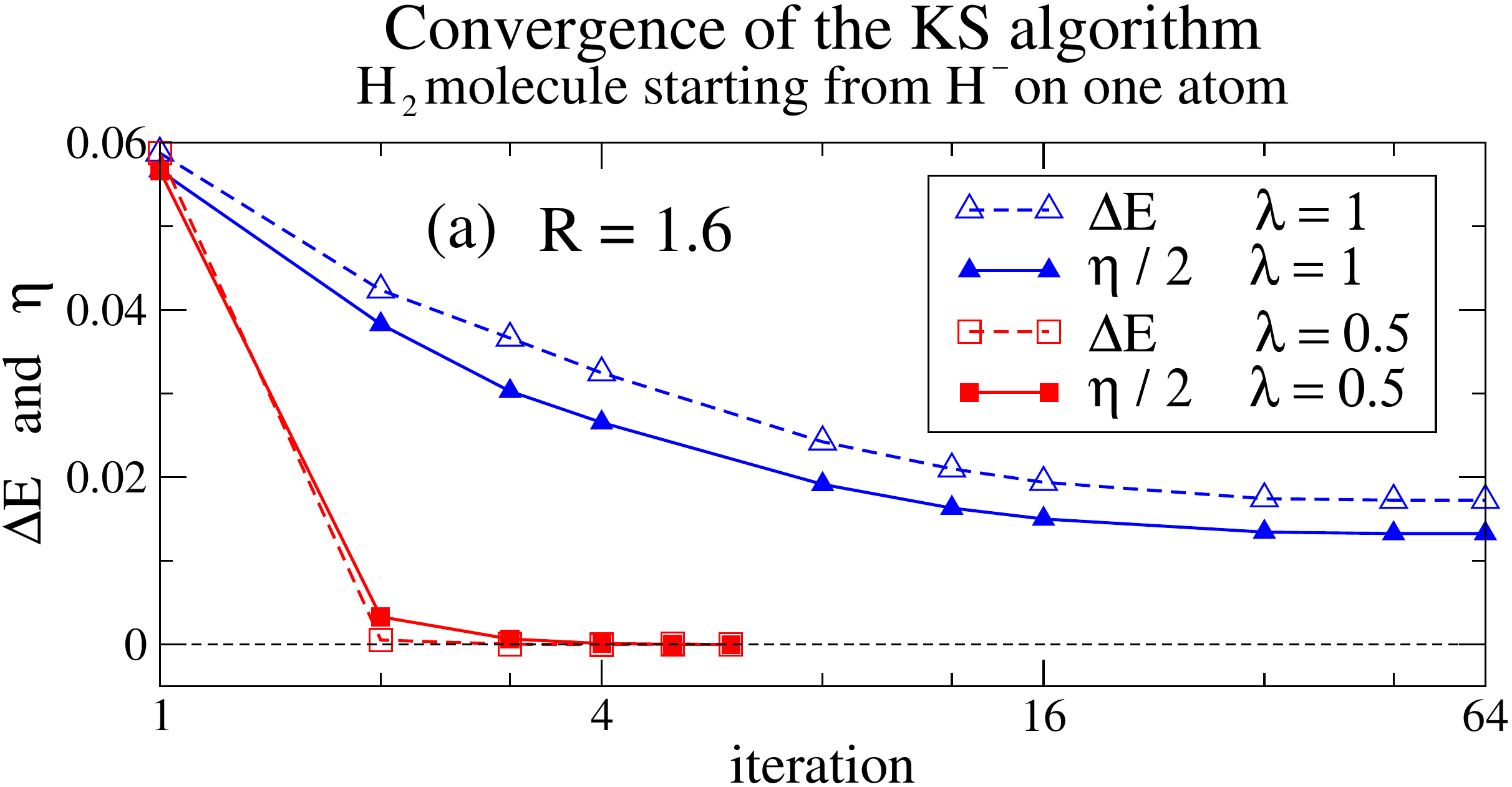}
\includegraphics[width=\columnwidth]{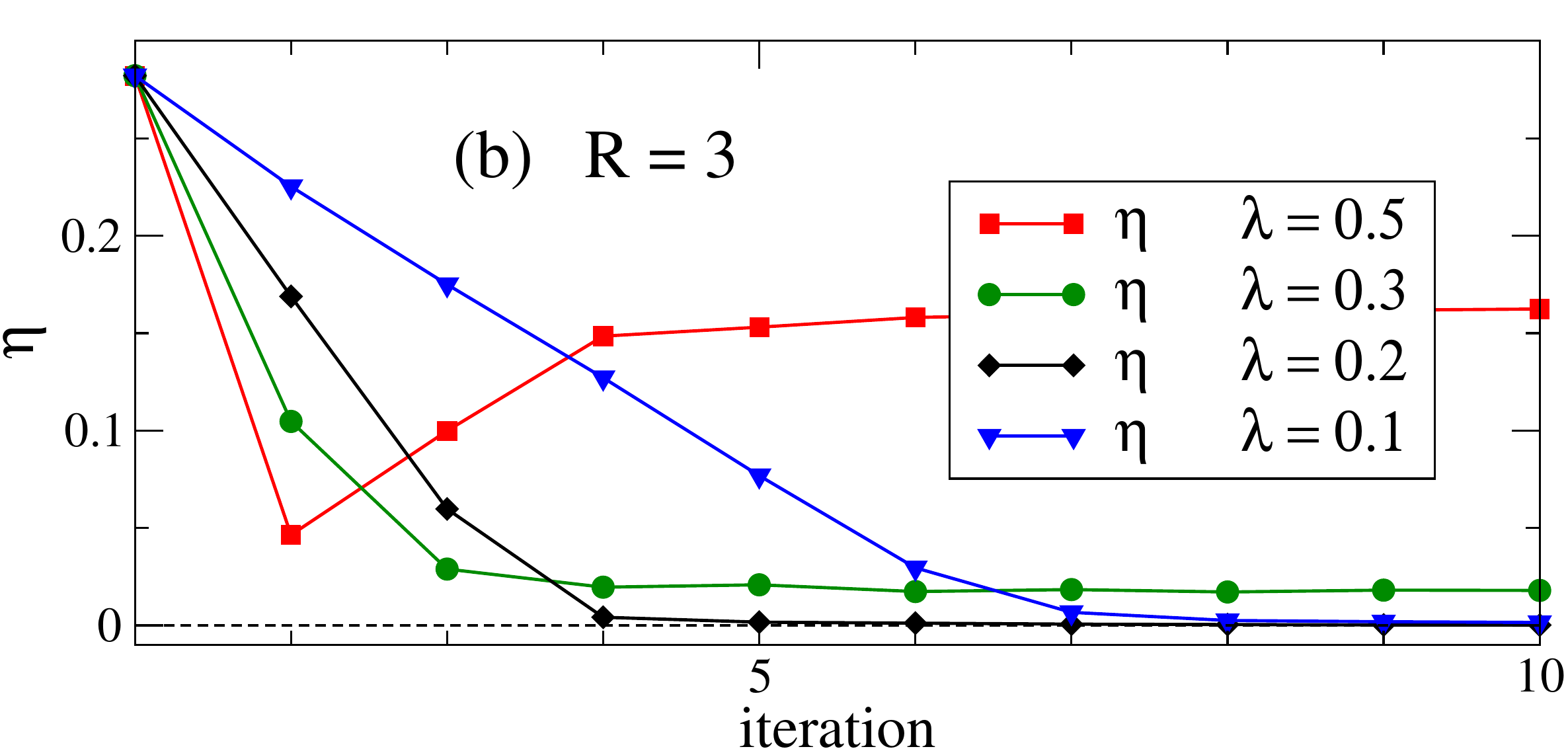}
\caption{
Differences in the density $\eta$ using \Eqref{Dn}
and the energy with $\Delta E = E_v[n'] - E\gs_v$,
for an H$_2$ molecule with (a) $R=1.6$ and (b) $R=3$.
In (b), the $\Delta E$ curves are omitted for clarity,
but are like those in (a).
}
\label{H2conv}
\vskip -0.5cm
\end{figure}

Consider the KS scheme applied to a
simple 1d H$_2$ molecule with bond length $R$ \cite{WSBW12}.  
Initialize the algorithm with
an asymmetric input density, an H$^-$ density centered on the left atom.  
Of course, no sensible KS calculation starts with such a density, but we do this to
amplify convergence issues.
In \Figref{H2conv}, we quantify the convergence of the KS algorithm
using $\eta$ from \Eqref{Dn} as well as energy differences from the ground-state.
For the equilibrium bond length ($R=1.6$), $\lambda$ may be chosen quite large ($\approx 0.5$);
but as the atoms are stretched to $R=3$, $\lambda$ must be $\lesssim 0.2$.
When $R=5$, even $\lambda = 0.01$ is too large to converge the calculation (not shown).
Thus, as the bond is stretched and the system develops strong static correlation \cite{WSBW12},
convergence becomes increasingly difficult.  
As more atoms are added to the chain (not shown), such as stretched H$_4$, even a reasonable
initial state converges very slowly.


\begin{figure}
\includegraphics[width=\columnwidth]{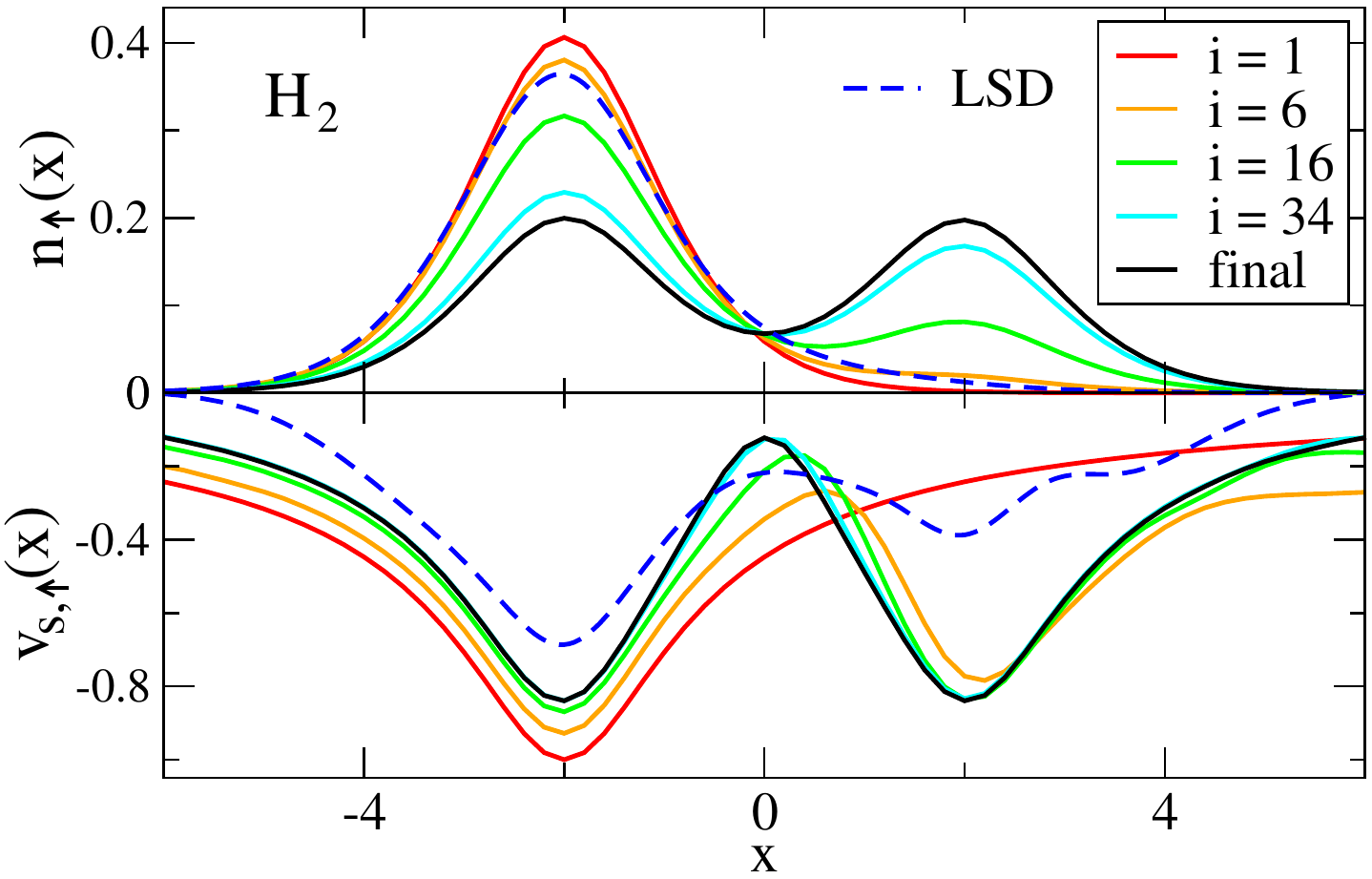}
\caption{
Starting an exact KS calculation of stretched H$_2$ with a spin-polarized 
density still converges (with $\lambda=0.5$) to the correct spin-singlet density.
For the same initial density,
the KS calculation with the local spin-density (LSD) approximation \cite{HFCV11}
converges to the broken spin-symmetry solution.
 }
\label{H2pol}
\vskip -0.5cm
\end{figure}

\prltitle{Consequences for real calculations}%
For approximate XC functionals, 
the corresponding $E_v[\n]$ is not,
in general, convex for every $v(\br)$, and
our corollaries do not
hold.  Consider H$_2$ in the local
spin-density approximation.   At and near equilibrium bond lengths, only one
stationary solution exists.  The approximate functional may or may not be convex.
But when the
bond is stretched beyond the infamous Coulson--Fischer point \cite{CF49,PSB95},
an unrestricted
solution of lower energy appears \cite{WSBW12}, as in \Figref{H2pol}, so    
the corresponding $E_v[\n]$ is not convex and
convergence with
our simple algorithm is not guaranteed.  
While the restricted solution is a saddle point,
the unrestricted solution is a local minimum.  Thus, only the
unrestricted solution behaves locally like the solution with the exact functional,
providing further rationale \cite{PSB95} for preferring such a solution over any restricted
one.
On the other hand, slowing of convergence as correlations become stronger is a real
effect, and not
an artifact of approximations.

We chose our simple algorithm to prove convergence, but 
many are more sophisticated and efficient
(see e.g.\ \cite{DS00,TOYJ04}).
Mixing KS potentials instead of densities
\cite{K81} can similarly be proven to converge, with
the advantage that 
all densities encountered are NI $v$-representable.

Finally, we expect that orbital degeneracies in 3d require the ensemble treatment \cite{V80,L82,L83,UK01}.
Further, extending the KS approach to use fractional occupation of electron orbitals  
(even in the case of non-degeneracy) 
may speed convergence \cite{RS99b} and allow KS-DFT
to more naturally handle strong static correlation \cite{GSB98b}.

The authors acknowledge DOE
grant  DE-SC0008696, and LW also thanks
the Korean global research network grant (No.~NRF-2010-220-C00017).
LW thanks Stefano Pittalis, Marcos Raydan, Robert
van Leeuwen and Mel Levy for helpful discussions.

\bibliography{ksprl}

\begin{thebibliography}{47}%
\makeatletter
\providecommand \@ifxundefined [1]{%
 \@ifx{#1\undefined}
}%
\providecommand \@ifnum [1]{%
 \ifnum #1\expandafter \@firstoftwo
 \else \expandafter \@secondoftwo
 \fi
}%
\providecommand \@ifx [1]{%
 \ifx #1\expandafter \@firstoftwo
 \else \expandafter \@secondoftwo
 \fi
}%
\providecommand \natexlab [1]{#1}%
\providecommand \enquote  [1]{``#1''}%
\providecommand \bibnamefont  [1]{#1}%
\providecommand \bibfnamefont [1]{#1}%
\providecommand \citenamefont [1]{#1}%
\providecommand \href@noop [0]{\@secondoftwo}%
\providecommand \href [0]{\begingroup \@sanitize@url \@href}%
\providecommand \@href[1]{\@@startlink{#1}\@@href}%
\providecommand \@@href[1]{\endgroup#1\@@endlink}%
\providecommand \@sanitize@url [0]{\catcode `\\12\catcode `\$12\catcode
  `\&12\catcode `\#12\catcode `\^12\catcode `\_12\catcode `\%12\relax}%
\providecommand \@@startlink[1]{}%
\providecommand \@@endlink[0]{}%
\providecommand \url  [0]{\begingroup\@sanitize@url \@url }%
\providecommand \@url [1]{\endgroup\@href {#1}{\urlprefix }}%
\providecommand \urlprefix  [0]{URL }%
\providecommand \Eprint [0]{\href }%
\providecommand \doibase [0]{http://dx.doi.org/}%
\providecommand \selectlanguage [0]{\@gobble}%
\providecommand \bibinfo  [0]{\@secondoftwo}%
\providecommand \bibfield  [0]{\@secondoftwo}%
\providecommand \translation [1]{[#1]}%
\providecommand \BibitemOpen [0]{}%
\providecommand \bibitemStop [0]{}%
\providecommand \bibitemNoStop [0]{.\EOS\space}%
\providecommand \EOS [0]{\spacefactor3000\relax}%
\providecommand \BibitemShut  [1]{\csname bibitem#1\endcsname}%
\let\auto@bib@innerbib\@empty
\bibitem [{\citenamefont {Kohn}\ and\ \citenamefont {Sham}(1965)}]{KS65}%
  \BibitemOpen
  \bibfield  {author} {\bibinfo {author} {\bibfnamefont {W.}~\bibnamefont
  {Kohn}}\ and\ \bibinfo {author} {\bibfnamefont {L.~J.}\ \bibnamefont
  {Sham}},\ }\bibfield  {title} {\enquote {\bibinfo {title} {Self-consistent
  equations including exchange and correlation effects},}\ }\href {\doibase
  10.1103/PhysRev.140.A1133} {\bibfield  {journal} {\bibinfo  {journal} {Phys.
  Rev.}\ }\textbf {\bibinfo {volume} {140}},\ \bibinfo {pages} {A1133--A1138}
  (\bibinfo {year} {1965})}\BibitemShut {NoStop}%
\bibitem [{\citenamefont {Burke}(2012)}]{B12}%
  \BibitemOpen
  \bibfield  {author} {\bibinfo {author} {\bibfnamefont {K.}~\bibnamefont
  {Burke}},\ }\bibfield  {title} {\enquote {\bibinfo {title} {Perspective on
  density functional theory},}\ }\href
  {http://link.aip.org/link/doi/10.1063/1.4704546} {\bibfield  {journal}
  {\bibinfo  {journal} {J. Chem. Phys.}\ }\textbf {\bibinfo {volume} {136}}
  (\bibinfo {year} {2012})}\BibitemShut {NoStop}%
\bibitem [{\citenamefont {Mori-S\'anchez}\ \emph {et~al.}(2009)\citenamefont
  {Mori-S\'anchez}, \citenamefont {Cohen},\ and\ \citenamefont {Yang}}]{MCY09}%
  \BibitemOpen
  \bibfield  {author} {\bibinfo {author} {\bibfnamefont {Paula}\ \bibnamefont
  {Mori-S\'anchez}}, \bibinfo {author} {\bibfnamefont {Aron~J.}\ \bibnamefont
  {Cohen}}, \ and\ \bibinfo {author} {\bibfnamefont {Weitao}\ \bibnamefont
  {Yang}},\ }\bibfield  {title} {\enquote {\bibinfo {title} {Discontinuous
  nature of the exchange-correlation functional in strongly correlated
  systems},}\ }\href {\doibase 10.1103/PhysRevLett.102.066403} {\bibfield
  {journal} {\bibinfo  {journal} {Phys. Rev. Lett.}\ }\textbf {\bibinfo
  {volume} {102}},\ \bibinfo {pages} {066403} (\bibinfo {year}
  {2009})}\BibitemShut {NoStop}%
\bibitem [{\citenamefont {Daniels}\ and\ \citenamefont
  {Scuseria}(2000)}]{DS00}%
  \BibitemOpen
  \bibfield  {author} {\bibinfo {author} {\bibfnamefont {Andrew~D.}\
  \bibnamefont {Daniels}}\ and\ \bibinfo {author} {\bibfnamefont {Gustavo~E.}\
  \bibnamefont {Scuseria}},\ }\bibfield  {title} {\enquote {\bibinfo {title}
  {Converging difficult scf cases with conjugate gradient density matrix
  search},}\ }\href {\doibase 10.1039/B000618L} {\bibfield  {journal} {\bibinfo
   {journal} {Phys. Chem. Chem. Phys.}\ }\textbf {\bibinfo {volume} {2}},\
  \bibinfo {pages} {2173--2176} (\bibinfo {year} {2000})}\BibitemShut {NoStop}%
\bibitem [{\citenamefont {Thogersen}\ \emph {et~al.}(2004)\citenamefont
  {Thogersen}, \citenamefont {Olsen}, \citenamefont {Yeager}, \citenamefont
  {Jorgensen}, \citenamefont {Salek},\ and\ \citenamefont {Helgaker}}]{TOYJ04}%
  \BibitemOpen
  \bibfield  {author} {\bibinfo {author} {\bibfnamefont {Lea}\ \bibnamefont
  {Thogersen}}, \bibinfo {author} {\bibfnamefont {Jeppe}\ \bibnamefont
  {Olsen}}, \bibinfo {author} {\bibfnamefont {Danny}\ \bibnamefont {Yeager}},
  \bibinfo {author} {\bibfnamefont {Poul}\ \bibnamefont {Jorgensen}}, \bibinfo
  {author} {\bibfnamefont {Pawel}\ \bibnamefont {Salek}}, \ and\ \bibinfo
  {author} {\bibfnamefont {Trygve}\ \bibnamefont {Helgaker}},\ }\bibfield
  {title} {\enquote {\bibinfo {title} {The trust-region self-consistent field
  method: Towards a black-box optimization in hartree--fock and kohn--sham
  theories},}\ }\href {\doibase 10.1063/1.1755673} {\bibfield  {journal}
  {\bibinfo  {journal} {The Journal of Chemical Physics}\ }\textbf {\bibinfo
  {volume} {121}},\ \bibinfo {pages} {16--27} (\bibinfo {year}
  {2004})}\BibitemShut {NoStop}%
\bibitem [{\citenamefont {Anisimov}\ \emph {et~al.}(1997)\citenamefont
  {Anisimov}, \citenamefont {Aryasetiawan},\ and\ \citenamefont
  {Lichtenstein}}]{AAL97}%
  \BibitemOpen
  \bibfield  {author} {\bibinfo {author} {\bibfnamefont {Vladimir~I}\
  \bibnamefont {Anisimov}}, \bibinfo {author} {\bibfnamefont {F}~\bibnamefont
  {Aryasetiawan}}, \ and\ \bibinfo {author} {\bibfnamefont {A~I}\ \bibnamefont
  {Lichtenstein}},\ }\bibfield  {title} {\enquote {\bibinfo {title}
  {First-principles calculations of the electronic structure and spectra of
  strongly correlated systems: the lda + u method},}\ }\href
  {http://stacks.iop.org/0953-8984/9/i=4/a=002} {\bibfield  {journal} {\bibinfo
   {journal} {Journal of Physics: Condensed Matter}\ }\textbf {\bibinfo
  {volume} {9}},\ \bibinfo {pages} {767} (\bibinfo {year} {1997})}\BibitemShut
  {NoStop}%
\bibitem [{\citenamefont {Heyd}\ \emph {et~al.}(2003)\citenamefont {Heyd},
  \citenamefont {Scuseria},\ and\ \citenamefont {Ernzerhof}}]{HSE06}%
  \BibitemOpen
  \bibfield  {author} {\bibinfo {author} {\bibfnamefont {Jochen}\ \bibnamefont
  {Heyd}}, \bibinfo {author} {\bibfnamefont {Gustavo~E.}\ \bibnamefont
  {Scuseria}}, \ and\ \bibinfo {author} {\bibfnamefont {Matthias}\ \bibnamefont
  {Ernzerhof}},\ }\bibfield  {title} {\enquote {\bibinfo {title} {Hybrid
  functionals based on a screened coulomb potential},}\ }\href {\doibase
  10.1063/1.1564060} {\bibfield  {journal} {\bibinfo  {journal} {The Journal of
  Chemical Physics}\ }\textbf {\bibinfo {volume} {118}},\ \bibinfo {pages}
  {8207--8215} (\bibinfo {year} {2003})},\ \bibinfo {note}
  {\href{http://link.aip.org/link/doi/10.1063/1.2204597}{{\em ibid.} {\bf 124},
  219906(E) (2006)}.}\BibitemShut {Stop}%
\bibitem [{\citenamefont {Assmann}\ \emph {et~al.}(2013)\citenamefont
  {Assmann}, \citenamefont {Blaha}, \citenamefont {Laskowski}, \citenamefont
  {Held}, \citenamefont {Okamoto},\ and\ \citenamefont {Sangiovanni}}]{ABLH13}%
  \BibitemOpen
  \bibfield  {author} {\bibinfo {author} {\bibfnamefont {Elias}\ \bibnamefont
  {Assmann}}, \bibinfo {author} {\bibfnamefont {Peter}\ \bibnamefont {Blaha}},
  \bibinfo {author} {\bibfnamefont {Robert}\ \bibnamefont {Laskowski}},
  \bibinfo {author} {\bibfnamefont {Karsten}\ \bibnamefont {Held}}, \bibinfo
  {author} {\bibfnamefont {Satoshi}\ \bibnamefont {Okamoto}}, \ and\ \bibinfo
  {author} {\bibfnamefont {Giorgio}\ \bibnamefont {Sangiovanni}},\ }\bibfield
  {title} {\enquote {\bibinfo {title} {Oxide heterostructures for efficient
  solar cells},}\ }\href {\doibase 10.1103/PhysRevLett.110.078701} {\bibfield
  {journal} {\bibinfo  {journal} {Phys. Rev. Lett.}\ }\textbf {\bibinfo
  {volume} {110}},\ \bibinfo {pages} {078701} (\bibinfo {year}
  {2013})}\BibitemShut {NoStop}%
\bibitem [{\citenamefont {Hohenberg}\ and\ \citenamefont {Kohn}(1964)}]{HK64}%
  \BibitemOpen
  \bibfield  {author} {\bibinfo {author} {\bibfnamefont {P.}~\bibnamefont
  {Hohenberg}}\ and\ \bibinfo {author} {\bibfnamefont {W.}~\bibnamefont
  {Kohn}},\ }\bibfield  {title} {\enquote {\bibinfo {title} {Inhomogeneous
  electron gas},}\ }\href {\doibase 10.1103/PhysRev.136.B864} {\bibfield
  {journal} {\bibinfo  {journal} {Phys. Rev.}\ }\textbf {\bibinfo {volume}
  {136}},\ \bibinfo {pages} {B864--B871} (\bibinfo {year} {1964})}\BibitemShut
  {NoStop}%
\bibitem [{\citenamefont {Dreizler}\ and\ \citenamefont {Gross}(1990)}]{DG90}%
  \BibitemOpen
  \bibfield  {author} {\bibinfo {author} {\bibfnamefont {R.~M.}\ \bibnamefont
  {Dreizler}}\ and\ \bibinfo {author} {\bibfnamefont {E.~K.~U.}\ \bibnamefont
  {Gross}},\ }\href@noop {} {\emph {\bibinfo {title} {Density Functional
  Theory: An Approach to the Quantum Many-Body Problem}}}\ (\bibinfo
  {publisher} {Springer--Verlag},\ \bibinfo {address} {Berlin},\ \bibinfo
  {year} {1990})\BibitemShut {NoStop}%
\bibitem [{\citenamefont {Stoudenmire}\ \emph {et~al.}(2012)\citenamefont
  {Stoudenmire}, \citenamefont {Wagner}, \citenamefont {White},\ and\
  \citenamefont {Burke}}]{SWWB12}%
  \BibitemOpen
  \bibfield  {author} {\bibinfo {author} {\bibfnamefont {E.~M.}\ \bibnamefont
  {Stoudenmire}}, \bibinfo {author} {\bibfnamefont {Lucas~O.}\ \bibnamefont
  {Wagner}}, \bibinfo {author} {\bibfnamefont {Steven~R.}\ \bibnamefont
  {White}}, \ and\ \bibinfo {author} {\bibfnamefont {Kieron}\ \bibnamefont
  {Burke}},\ }\bibfield  {title} {\enquote {\bibinfo {title} {One-dimensional
  continuum electronic structure with the density-matrix renormalization group
  and its implications for density-functional theory},}\ }\href {\doibase
  10.1103/PhysRevLett.109.056402} {\bibfield  {journal} {\bibinfo  {journal}
  {Phys. Rev. Lett.}\ }\textbf {\bibinfo {volume} {109}},\ \bibinfo {pages}
  {056402} (\bibinfo {year} {2012})}\BibitemShut {NoStop}%
\bibitem [{\citenamefont {Wagner}\ \emph {et~al.}(2012)\citenamefont {Wagner},
  \citenamefont {Stoudenmire}, \citenamefont {Burke},\ and\ \citenamefont
  {White}}]{WSBW12}%
  \BibitemOpen
  \bibfield  {author} {\bibinfo {author} {\bibfnamefont {Lucas~O.}\
  \bibnamefont {Wagner}}, \bibinfo {author} {\bibfnamefont {E.M.}\ \bibnamefont
  {Stoudenmire}}, \bibinfo {author} {\bibfnamefont {Kieron}\ \bibnamefont
  {Burke}}, \ and\ \bibinfo {author} {\bibfnamefont {Steven~R.}\ \bibnamefont
  {White}},\ }\bibfield  {title} {\enquote {\bibinfo {title} {Reference
  electronic structure calculations in one dimension},}\ }\href {\doibase
  DOI:10.1039/C2CP24118H} {\bibfield  {journal} {\bibinfo  {journal} {Phys.
  Chem. Chem. Phys.}\ }\textbf {\bibinfo {volume} {14}},\ \bibinfo {pages}
  {8581 -- 8590} (\bibinfo {year} {2012})}\BibitemShut {NoStop}%
\bibitem [{\citenamefont {Chayes}\ \emph {et~al.}(1985)\citenamefont {Chayes},
  \citenamefont {Chayes},\ and\ \citenamefont {Ruskai}}]{CCR85}%
  \BibitemOpen
  \bibfield  {author} {\bibinfo {author} {\bibfnamefont {J.T.}\ \bibnamefont
  {Chayes}}, \bibinfo {author} {\bibfnamefont {L.}~\bibnamefont {Chayes}}, \
  and\ \bibinfo {author} {\bibfnamefont {MaryBeth}\ \bibnamefont {Ruskai}},\
  }\bibfield  {title} {\enquote {\bibinfo {title} {Density functional approach
  to quantum lattice systems},}\ }\href {\doibase 10.1007/BF01010474}
  {\bibfield  {journal} {\bibinfo  {journal} {Journal of Statistical Physics}\
  }\textbf {\bibinfo {volume} {38}},\ \bibinfo {pages} {497--518} (\bibinfo
  {year} {1985})}\BibitemShut {NoStop}%
\bibitem [{\citenamefont {Levy}\ and\ \citenamefont {Perdew}(1985)}]{LP85}%
  \BibitemOpen
  \bibfield  {author} {\bibinfo {author} {\bibfnamefont {M.}~\bibnamefont
  {Levy}}\ and\ \bibinfo {author} {\bibfnamefont {J.P.}\ \bibnamefont
  {Perdew}},\ }\bibfield  {title} {\enquote {\bibinfo {title}
  {Hellmann-feynman, virial, and scaling requisites for the exact universal
  density functionals. shape of the correlation potential and diamagnetic
  susceptibility for atoms},}\ }\href {\doibase 10.1103/PhysRevA.32.2010}
  {\bibfield  {journal} {\bibinfo  {journal} {Phys. Rev. A}\ }\textbf {\bibinfo
  {volume} {32}},\ \bibinfo {pages} {2010} (\bibinfo {year}
  {1985})}\BibitemShut {NoStop}%
\bibitem [{\citenamefont {Perdew}\ \emph {et~al.}(1996)\citenamefont {Perdew},
  \citenamefont {Burke},\ and\ \citenamefont {Ernzerhof}}]{PBE96}%
  \BibitemOpen
  \bibfield  {author} {\bibinfo {author} {\bibfnamefont {John~P.}\ \bibnamefont
  {Perdew}}, \bibinfo {author} {\bibfnamefont {Kieron}\ \bibnamefont {Burke}},
  \ and\ \bibinfo {author} {\bibfnamefont {Matthias}\ \bibnamefont
  {Ernzerhof}},\ }\bibfield  {title} {\enquote {\bibinfo {title} {Generalized
  gradient approximation made simple},}\ }\href {\doibase
  10.1103/PhysRevLett.77.3865} {\bibfield  {journal} {\bibinfo  {journal}
  {Phys. Rev. Lett.}\ }\textbf {\bibinfo {volume} {77}},\ \bibinfo {pages}
  {3865--3868} (\bibinfo {year} {1996})},\ \bibinfo {note} {{\it ibid.} {\bf
  78}, 1396(E) (1997)}\BibitemShut {NoStop}%
\bibitem [{\citenamefont {Perdew}\ and\ \citenamefont {Kurth}(2003)}]{PK03}%
  \BibitemOpen
  \bibfield  {author} {\bibinfo {author} {\bibfnamefont {John~P.}\ \bibnamefont
  {Perdew}}\ and\ \bibinfo {author} {\bibfnamefont {Stefan}\ \bibnamefont
  {Kurth}},\ }\enquote {\bibinfo {title} {Density functionals for
  non-relativistic {C}oulomb systems in the new century},}\ in\ \href {\doibase
  10.1007/3-540-37072-2_1} {\emph {\bibinfo {booktitle} {A Primer in Density
  Functional Theory}}},\ \bibinfo {editor} {edited by\ \bibinfo {editor}
  {\bibfnamefont {Carlos}\ \bibnamefont {Fiolhais}}, \bibinfo {editor}
  {\bibfnamefont {Fernando}\ \bibnamefont {Nogueira}}, \ and\ \bibinfo {editor}
  {\bibfnamefont {Miguel A.~L.}\ \bibnamefont {Marques}}}\ (\bibinfo
  {publisher} {Springer},\ \bibinfo {address} {Berlin / Heidelberg},\ \bibinfo
  {year} {2003})\ pp.\ \bibinfo {pages} {1--55}\BibitemShut {NoStop}%
\bibitem [{\citenamefont {Perdew}\ \emph {et~al.}(2005)\citenamefont {Perdew},
  \citenamefont {Ruzsinszky}, \citenamefont {Tao}, \citenamefont {Staroverov},
  \citenamefont {Scuseria},\ and\ \citenamefont {Csonka}}]{PRTS05}%
  \BibitemOpen
  \bibfield  {author} {\bibinfo {author} {\bibfnamefont {John~P.}\ \bibnamefont
  {Perdew}}, \bibinfo {author} {\bibfnamefont {Adrienn}\ \bibnamefont
  {Ruzsinszky}}, \bibinfo {author} {\bibfnamefont {Jianmin}\ \bibnamefont
  {Tao}}, \bibinfo {author} {\bibfnamefont {Viktor~N.}\ \bibnamefont
  {Staroverov}}, \bibinfo {author} {\bibfnamefont {Gustavo~E.}\ \bibnamefont
  {Scuseria}}, \ and\ \bibinfo {author} {\bibfnamefont {Gabor~I.}\ \bibnamefont
  {Csonka}},\ }\bibfield  {title} {\enquote {\bibinfo {title} {Prescription for
  the design and selection of density functional approximations: More
  constraint satisfaction with fewer fits},}\ }\href {\doibase
  10.1063/1.1904565} {\bibfield  {journal} {\bibinfo  {journal} {The Journal of
  Chemical Physics}\ }\textbf {\bibinfo {volume} {123}},\ \bibinfo {eid}
  {062201} (\bibinfo {year} {2005})}\BibitemShut {NoStop}%
\bibitem [{\citenamefont {Perdew}\ \emph {et~al.}(1982)\citenamefont {Perdew},
  \citenamefont {Parr}, \citenamefont {Levy},\ and\ \citenamefont
  {Balduz}}]{PPLB82}%
  \BibitemOpen
  \bibfield  {author} {\bibinfo {author} {\bibfnamefont {John~P.}\ \bibnamefont
  {Perdew}}, \bibinfo {author} {\bibfnamefont {Robert~G.}\ \bibnamefont
  {Parr}}, \bibinfo {author} {\bibfnamefont {Mel}\ \bibnamefont {Levy}}, \ and\
  \bibinfo {author} {\bibfnamefont {Jose~L.}\ \bibnamefont {Balduz}},\
  }\bibfield  {title} {\enquote {\bibinfo {title} {Density-functional theory
  for fractional particle number: Derivative discontinuities of the energy},}\
  }\href {\doibase 10.1103/PhysRevLett.49.1691} {\bibfield  {journal} {\bibinfo
   {journal} {Phys. Rev. Lett.}\ }\textbf {\bibinfo {volume} {49}},\ \bibinfo
  {pages} {1691--1694} (\bibinfo {year} {1982})}\BibitemShut {NoStop}%
\bibitem [{\citenamefont {Anantharaman}\ and\ \citenamefont
  {Canc\`es}(2009)}]{AC09}%
  \BibitemOpen
  \bibfield  {author} {\bibinfo {author} {\bibfnamefont {Arnaud}\ \bibnamefont
  {Anantharaman}}\ and\ \bibinfo {author} {\bibfnamefont {Eric}\ \bibnamefont
  {Canc\`es}},\ }\bibfield  {title} {\enquote {\bibinfo {title} {Existence of
  minimizers for kohn–sham models in quantum chemistry},}\ }\href {\doibase
  http://dx.doi.org/10.1016/j.anihpc.2009.06.003} {\bibfield  {journal}
  {\bibinfo  {journal} {Annales de l'Institut Henri Poincare (C) Non Linear
  Analysis}\ }\textbf {\bibinfo {volume} {26}},\ \bibinfo {pages} {2425 --
  2455} (\bibinfo {year} {2009})}\BibitemShut {NoStop}%
\bibitem [{\citenamefont {Canc\`es}\ and\ \citenamefont
  {Le~Bris}(2000)}]{CL00}%
  \BibitemOpen
  \bibfield  {author} {\bibinfo {author} {\bibfnamefont {Eric}\ \bibnamefont
  {Canc\`es}}\ and\ \bibinfo {author} {\bibfnamefont {Claude}\ \bibnamefont
  {Le~Bris}},\ }\bibfield  {title} {\enquote {\bibinfo {title} {On the
  convergence of scf algorithms for the hartree-fock equations},}\ }\href
  {\doibase 10.1051/m2an:2000102} {\bibfield  {journal} {\bibinfo  {journal}
  {ESAIM: Mathematical Modelling and Numerical Analysis}\ }\textbf {\bibinfo
  {volume} {34}},\ \bibinfo {pages} {749--774} (\bibinfo {year}
  {2000})}\BibitemShut {NoStop}%
\bibitem [{Note1()}]{Note1}%
  \BibitemOpen
  \bibinfo {note} {The precise restrictions on potentials are detailed in
  Ref.~\cite {L83}; Coulomb potentials are allowed.}\BibitemShut {Stop}%
\bibitem [{\citenamefont {Lieb}(1983)}]{L83}%
  \BibitemOpen
  \bibfield  {author} {\bibinfo {author} {\bibfnamefont {Elliott~H.}\
  \bibnamefont {Lieb}},\ }\bibfield  {title} {\enquote {\bibinfo {title}
  {Density functionals for coulomb systems},}\ }\href {\doibase
  10.1002/qua.560240302} {\bibfield  {journal} {\bibinfo  {journal} {Int. J.
  Quantum Chem.}\ }\textbf {\bibinfo {volume} {24}},\ \bibinfo {pages}
  {243--277} (\bibinfo {year} {1983})}\BibitemShut {NoStop}%
\bibitem [{\citenamefont {Fiolhais}\ \emph {et~al.}(2003)\citenamefont
  {Fiolhais}, \citenamefont {Nogueira},\ and\ \citenamefont {Marques}}]{FNM03}%
  \BibitemOpen
  \bibfield  {author} {\bibinfo {author} {\bibfnamefont {Carlos}\ \bibnamefont
  {Fiolhais}}, \bibinfo {author} {\bibfnamefont {F.}~\bibnamefont {Nogueira}},
  \ and\ \bibinfo {author} {\bibfnamefont {M.}~\bibnamefont {Marques}},\
  }\href@noop {} {\emph {\bibinfo {title} {A Primer in Density Functional
  Theory}}}\ (\bibinfo  {publisher} {Springer-Verlag},\ \bibinfo {address} {New
  York},\ \bibinfo {year} {2003})\BibitemShut {NoStop}%
\bibitem [{Note2()}]{Note2}%
  \BibitemOpen
  \bibinfo {note} {\label {orbrot}An orbital rotation among degenerate orbitals
  may also be required; see Ref.~\cite {UK01}.}\BibitemShut {Stop}%
\bibitem [{\citenamefont {Ullrich}\ and\ \citenamefont {Kohn}(2001)}]{UK01}%
  \BibitemOpen
  \bibfield  {author} {\bibinfo {author} {\bibfnamefont {C.~A.}\ \bibnamefont
  {Ullrich}}\ and\ \bibinfo {author} {\bibfnamefont {W.}~\bibnamefont {Kohn}},\
  }\bibfield  {title} {\enquote {\bibinfo {title} {Kohn-sham theory for
  ground-state ensembles},}\ }\href {\doibase 10.1103/PhysRevLett.87.093001}
  {\bibfield  {journal} {\bibinfo  {journal} {Phys. Rev. Lett.}\ }\textbf
  {\bibinfo {volume} {87}},\ \bibinfo {pages} {093001} (\bibinfo {year}
  {2001})}\BibitemShut {NoStop}%
\bibitem [{\citenamefont {Levy}(1982)}]{L82}%
  \BibitemOpen
  \bibfield  {author} {\bibinfo {author} {\bibfnamefont {Mel}\ \bibnamefont
  {Levy}},\ }\bibfield  {title} {\enquote {\bibinfo {title} {Electron densities
  in search of hamiltonians},}\ }\href {\doibase 10.1103/PhysRevA.26.1200}
  {\bibfield  {journal} {\bibinfo  {journal} {Phys. Rev. A}\ }\textbf {\bibinfo
  {volume} {26}},\ \bibinfo {pages} {1200--1208} (\bibinfo {year}
  {1982})}\BibitemShut {NoStop}%
\bibitem [{\citenamefont {van Leeuwen}(2003)}]{Leeuwen}%
  \BibitemOpen
  \bibfield  {author} {\bibinfo {author} {\bibfnamefont {Robert}\ \bibnamefont
  {van Leeuwen}},\ }\bibfield  {title} {\enquote {\bibinfo {title} {Density
  functional approach to the many-body problem: Key concepts and exact
  functionals},}\ \ }(\bibinfo  {publisher} {Academic Press},\ \bibinfo {year}
  {2003})\ pp.\ \bibinfo {pages} {25 -- 94}\BibitemShut {NoStop}%
\bibitem [{Note3()}]{Note3}%
  \BibitemOpen
  \bibinfo {note} {We restrict our attention to finite systems, since in an
  extended system Eq.~\protect \textup {\hbox {\mathsurround \z@ \protect
  \normalfont (\ignorespaces \ref {ineq}\unskip \@@italiccorr )}} would be ill
  defined.}\BibitemShut {Stop}%
\bibitem [{\citenamefont {Birgin}\ \emph {et~al.}(2003)\citenamefont {Birgin},
  \citenamefont {Martínez},\ and\ \citenamefont {Raydan}}]{BMR03}%
  \BibitemOpen
  \bibfield  {author} {\bibinfo {author} {\bibfnamefont {Ernesto~G.}\
  \bibnamefont {Birgin}}, \bibinfo {author} {\bibfnamefont {José~Mario}\
  \bibnamefont {Martínez}}, \ and\ \bibinfo {author} {\bibfnamefont {Marcos}\
  \bibnamefont {Raydan}},\ }\bibfield  {title} {\enquote {\bibinfo {title}
  {Inexact spectral projected gradient methods on convex sets},}\ }\href
  {\doibase 10.1093/imanum/23.4.539} {\bibfield  {journal} {\bibinfo  {journal}
  {IMA Journal of Numerical Analysis}\ }\textbf {\bibinfo {volume} {23}},\
  \bibinfo {pages} {539--559} (\bibinfo {year} {2003})}\BibitemShut {NoStop}%
\bibitem [{Note4()}]{Note4}%
  \BibitemOpen
  \bibinfo {note} {With respect to gradient descent minimization, we assume
  that functionals behave like their less dimensionful counterparts, functions.
  A more comprehensive analysis is beyond the scope of this work.}\BibitemShut
  {Stop}%
\bibitem [{Note5()}]{Note5}%
  \BibitemOpen
  \bibinfo {note} {In the language of DFT, we can find an
  ensemble-$v$-representable density which is arbitarily close to a density
  $\protect \mathaccentV {bar}016{n}({\protect \bf r})$ which has reasonable
  properties: integrating to a finite particle number $N$, with a finite von
  Weizs{\"a}cker kinetic energy, i.e.\ $\protect \sqrt {\protect \mathaccentV
  {bar}016{n}({\protect \bf r})} \in H^1(\protect \mathbb {R}^3)$, and being
  nowhere negative \cite {L83,Leeuwen}. This set of reasonable densities is
  convex \cite {Leeuwen}; so that $n_\lambda ({\protect \bf r})$ is reasonable
  if both $n({\protect \bf r})$ and $n'({\protect \bf r})$ are, which is the
  case. The examples of Ref.~\cite {EE83} are not in this set.}\BibitemShut
  {Stop}%
\bibitem [{\citenamefont {Englisch}\ and\ \citenamefont
  {Englisch}(1983)}]{EE83}%
  \BibitemOpen
  \bibfield  {author} {\bibinfo {author} {\bibfnamefont {H.}~\bibnamefont
  {Englisch}}\ and\ \bibinfo {author} {\bibfnamefont {R.}~\bibnamefont
  {Englisch}},\ }\bibfield  {title} {\enquote {\bibinfo {title}
  {Hohenberg-{K}ohn theorem and non-{V}-representable densities},}\ }\href
  {\doibase 10.1016/0378-4371(83)90254-6} {\bibfield  {journal} {\bibinfo
  {journal} {Physica A: Statistical Mechanics and its Applications}\ }\textbf
  {\bibinfo {volume} {121}},\ \bibinfo {pages} {253 -- 268} (\bibinfo {year}
  {1983})}\BibitemShut {NoStop}%
\bibitem [{\citenamefont {Valone}(1980)}]{V80}%
  \BibitemOpen
  \bibfield  {author} {\bibinfo {author} {\bibfnamefont {Steven~M.}\
  \bibnamefont {Valone}},\ }\bibfield  {title} {\enquote {\bibinfo {title} {A
  one-to-one mapping between one-particle densities and some n-particle
  ensembles},}\ }\href {\doibase 10.1063/1.440656} {\bibfield  {journal}
  {\bibinfo  {journal} {The Journal of Chemical Physics}\ }\textbf {\bibinfo
  {volume} {73}},\ \bibinfo {pages} {4653--4655} (\bibinfo {year}
  {1980})}\BibitemShut {NoStop}%
\bibitem [{\citenamefont {Perdew}\ and\ \citenamefont {Levy}(1985)}]{PL85}%
  \BibitemOpen
  \bibfield  {author} {\bibinfo {author} {\bibfnamefont {John~P.}\ \bibnamefont
  {Perdew}}\ and\ \bibinfo {author} {\bibfnamefont {Mel}\ \bibnamefont
  {Levy}},\ }\bibfield  {title} {\enquote {\bibinfo {title} {Extrema of the
  density functional for the energy: Excited states from the ground-state
  theory},}\ }\href {\doibase 10.1103/PhysRevB.31.6264} {\bibfield  {journal}
  {\bibinfo  {journal} {Phys. Rev. B}\ }\textbf {\bibinfo {volume} {31}},\
  \bibinfo {pages} {6264--6272} (\bibinfo {year} {1985})}\BibitemShut {NoStop}%
\bibitem [{\citenamefont {Levy}(1979)}]{L79}%
  \BibitemOpen
  \bibfield  {author} {\bibinfo {author} {\bibfnamefont {Mel}\ \bibnamefont
  {Levy}},\ }\bibfield  {title} {\enquote {\bibinfo {title} {Universal
  variational functionals of electron densities, first-order density matrices,
  and natural spin-orbitals and solution of the $v$-representability
  problem},}\ }\href {http://www.pnas.org/content/76/12/6062.abstract}
  {\bibfield  {journal} {\bibinfo  {journal} {Proceedings of the National
  Academy of Sciences of the United States of America}\ }\textbf {\bibinfo
  {volume} {76}},\ \bibinfo {pages} {6062--6065} (\bibinfo {year}
  {1979})}\BibitemShut {NoStop}%
\bibitem [{\citenamefont {Schuch}\ and\ \citenamefont
  {Verstraete}(2009)}]{SV09}%
  \BibitemOpen
  \bibfield  {author} {\bibinfo {author} {\bibfnamefont {Norbert}\ \bibnamefont
  {Schuch}}\ and\ \bibinfo {author} {\bibfnamefont {Frank}\ \bibnamefont
  {Verstraete}},\ }\bibfield  {title} {\enquote {\bibinfo {title}
  {Computational complexity of interacting electrons and fundamental
  limitations of density functional theory},}\ }\href {\doibase
  10.1038/nphys1370} {\bibfield  {journal} {\bibinfo  {journal} {Nat.\ Phys.}\
  }\textbf {\bibinfo {volume} {5}},\ \bibinfo {pages} {732--735} (\bibinfo
  {year} {2009})}\BibitemShut {NoStop}%
\bibitem [{\citenamefont {Broyden}(1965)}]{B65}%
  \BibitemOpen
  \bibfield  {author} {\bibinfo {author} {\bibfnamefont {C.G.}\ \bibnamefont
  {Broyden}},\ }\bibfield  {title} {\enquote {\bibinfo {title} {A class of
  methods for solving nonlinear simultaneous equations},}\ }\href {\doibase
  10.1090/S0025-5718-1965-0198670-6} {\bibfield  {journal} {\bibinfo  {journal}
  {Mathematics of Computation}\ }\textbf {\bibinfo {volume} {19}},\ \bibinfo
  {pages} {577--593} (\bibinfo {year} {1965})}\BibitemShut {NoStop}%
\bibitem [{\citenamefont {White}(1992)}]{White:1992}%
  \BibitemOpen
  \bibfield  {author} {\bibinfo {author} {\bibfnamefont {Steven~R.}\
  \bibnamefont {White}},\ }\bibfield  {title} {\enquote {\bibinfo {title}
  {Density matrix formulation for quantum renormalization groups},}\ }\href
  {http://link.aps.org/doi/10.1103/PhysRevLett.69.2863} {\bibfield  {journal}
  {\bibinfo  {journal} {Phys. Rev. Lett.}\ }\textbf {\bibinfo {volume} {69}},\
  \bibinfo {pages} {2863} (\bibinfo {year} {1992})}\BibitemShut {NoStop}%
\bibitem [{\citenamefont {White}(1993)}]{White:1993a}%
  \BibitemOpen
  \bibfield  {author} {\bibinfo {author} {\bibfnamefont {Steven~R.}\
  \bibnamefont {White}},\ }\bibfield  {title} {\enquote {\bibinfo {title}
  {Density-matrix algorithms for quantum renormalization groups},}\ }\href
  {http://link.aps.org/doi/10.1103/PhysRevB.48.10345} {\bibfield  {journal}
  {\bibinfo  {journal} {Phys. Rev. B}\ }\textbf {\bibinfo {volume} {48}},\
  \bibinfo {pages} {10345} (\bibinfo {year} {1993})}\BibitemShut {NoStop}%
\bibitem [{\citenamefont {Thiele}\ \emph {et~al.}(2008)\citenamefont {Thiele},
  \citenamefont {Gross},\ and\ \citenamefont {K\"ummel}}]{TGK08}%
  \BibitemOpen
  \bibfield  {author} {\bibinfo {author} {\bibfnamefont {M.}~\bibnamefont
  {Thiele}}, \bibinfo {author} {\bibfnamefont {E.~K.~U.}\ \bibnamefont
  {Gross}}, \ and\ \bibinfo {author} {\bibfnamefont {S.}~\bibnamefont
  {K\"ummel}},\ }\bibfield  {title} {\enquote {\bibinfo {title} {Adiabatic
  approximation in nonperturbative time-dependent density-functional theory},}\
  }\href {\doibase 10.1103/PhysRevLett.100.153004} {\bibfield  {journal}
  {\bibinfo  {journal} {Phys. Rev. Lett.}\ }\textbf {\bibinfo {volume} {100}},\
  \bibinfo {pages} {153004} (\bibinfo {year} {2008})}\BibitemShut {NoStop}%
\bibitem [{\citenamefont {Coe}\ \emph {et~al.}(2009)\citenamefont {Coe},
  \citenamefont {Capelle},\ and\ \citenamefont {D'Amico}}]{CCD09}%
  \BibitemOpen
  \bibfield  {author} {\bibinfo {author} {\bibfnamefont {J.~P.}\ \bibnamefont
  {Coe}}, \bibinfo {author} {\bibfnamefont {K.}~\bibnamefont {Capelle}}, \ and\
  \bibinfo {author} {\bibfnamefont {I.}~\bibnamefont {D'Amico}},\ }\bibfield
  {title} {\enquote {\bibinfo {title} {Reverse engineering in many-body quantum
  physics: Correspondence between many-body systems and effective
  single-particle equations},}\ }\href {\doibase 10.1103/PhysRevA.79.032504}
  {\bibfield  {journal} {\bibinfo  {journal} {Phys. Rev. A}\ }\textbf {\bibinfo
  {volume} {79}},\ \bibinfo {pages} {032504} (\bibinfo {year}
  {2009})}\BibitemShut {NoStop}%
\bibitem [{\citenamefont {Helbig}\ \emph {et~al.}(2011)\citenamefont {Helbig},
  \citenamefont {Fuks}, \citenamefont {Casula}, \citenamefont {Verstraete},
  \citenamefont {Marques}, \citenamefont {Tokatly},\ and\ \citenamefont
  {Rubio}}]{HFCV11}%
  \BibitemOpen
  \bibfield  {author} {\bibinfo {author} {\bibfnamefont {N.}~\bibnamefont
  {Helbig}}, \bibinfo {author} {\bibfnamefont {J.~I.}\ \bibnamefont {Fuks}},
  \bibinfo {author} {\bibfnamefont {M.}~\bibnamefont {Casula}}, \bibinfo
  {author} {\bibfnamefont {M.~J.}\ \bibnamefont {Verstraete}}, \bibinfo
  {author} {\bibfnamefont {M.~A.~L.}\ \bibnamefont {Marques}}, \bibinfo
  {author} {\bibfnamefont {I.~V.}\ \bibnamefont {Tokatly}}, \ and\ \bibinfo
  {author} {\bibfnamefont {A.}~\bibnamefont {Rubio}},\ }\bibfield  {title}
  {\enquote {\bibinfo {title} {Density functional theory beyond the linear
  regime: Validating an adiabatic local density approximation},}\ }\href
  {\doibase 10.1103/PhysRevA.83.032503} {\bibfield  {journal} {\bibinfo
  {journal} {Phys. Rev. A}\ }\textbf {\bibinfo {volume} {83}},\ \bibinfo
  {pages} {032503} (\bibinfo {year} {2011})}\BibitemShut {NoStop}%
\bibitem [{\citenamefont {Coulson}\ and\ \citenamefont {Fischer}(1949)}]{CF49}%
  \BibitemOpen
  \bibfield  {author} {\bibinfo {author} {\bibfnamefont {C.A.}\ \bibnamefont
  {Coulson}}\ and\ \bibinfo {author} {\bibfnamefont {I.}~\bibnamefont
  {Fischer}},\ }\bibfield  {title} {\enquote {\bibinfo {title} {Xxxiv. notes on
  the molecular orbital treatment of the hydrogen molecule},}\ }\href {\doibase
  10.1080/14786444908521726} {\bibfield  {journal} {\bibinfo  {journal}
  {Philosophical Magazine Series 7}\ }\textbf {\bibinfo {volume} {40}},\
  \bibinfo {pages} {386--393} (\bibinfo {year} {1949})}\BibitemShut {NoStop}%
\bibitem [{\citenamefont {Perdew}\ \emph {et~al.}(1995)\citenamefont {Perdew},
  \citenamefont {Savin},\ and\ \citenamefont {Burke}}]{PSB95}%
  \BibitemOpen
  \bibfield  {author} {\bibinfo {author} {\bibfnamefont {John~P.}\ \bibnamefont
  {Perdew}}, \bibinfo {author} {\bibfnamefont {Andreas}\ \bibnamefont {Savin}},
  \ and\ \bibinfo {author} {\bibfnamefont {Kieron}\ \bibnamefont {Burke}},\
  }\bibfield  {title} {\enquote {\bibinfo {title} {Escaping the symmetry
  dilemma through a pair-density interpretation of spin-density functional
  theory},}\ }\href {\doibase 10.1103/PhysRevA.51.4531} {\bibfield  {journal}
  {\bibinfo  {journal} {Phys. Rev. A}\ }\textbf {\bibinfo {volume} {51}},\
  \bibinfo {pages} {4531--4541} (\bibinfo {year} {1995})}\BibitemShut {NoStop}%
\bibitem [{\citenamefont {Kerker}(1981)}]{K81}%
  \BibitemOpen
  \bibfield  {author} {\bibinfo {author} {\bibfnamefont {G.~P.}\ \bibnamefont
  {Kerker}},\ }\bibfield  {title} {\enquote {\bibinfo {title} {Efficient
  iteration scheme for self-consistent pseudopotential calculations},}\ }\href
  {\doibase 10.1103/PhysRevB.23.3082} {\bibfield  {journal} {\bibinfo
  {journal} {Phys. Rev. B}\ }\textbf {\bibinfo {volume} {23}},\ \bibinfo
  {pages} {3082--3084} (\bibinfo {year} {1981})}\BibitemShut {NoStop}%
\bibitem [{\citenamefont {Rabuck}\ and\ \citenamefont
  {Scuseria}(1999)}]{RS99b}%
  \BibitemOpen
  \bibfield  {author} {\bibinfo {author} {\bibfnamefont {Angela~D.}\
  \bibnamefont {Rabuck}}\ and\ \bibinfo {author} {\bibfnamefont {Gustavo~E.}\
  \bibnamefont {Scuseria}},\ }\bibfield  {title} {\enquote {\bibinfo {title}
  {Improving self-consistent field convergence by varying occupation
  numbers},}\ }\href {\doibase 10.1063/1.478177} {\bibfield  {journal}
  {\bibinfo  {journal} {The Journal of Chemical Physics}\ }\textbf {\bibinfo
  {volume} {110}},\ \bibinfo {pages} {695--700} (\bibinfo {year}
  {1999})}\BibitemShut {NoStop}%
\bibitem [{\citenamefont {Schipper}\ \emph {et~al.}(1998)\citenamefont
  {Schipper}, \citenamefont {Gritsenko},\ and\ \citenamefont
  {Baerends}}]{GSB98b}%
  \BibitemOpen
  \bibfield  {author} {\bibinfo {author} {\bibfnamefont {P.~R.~T.}\
  \bibnamefont {Schipper}}, \bibinfo {author} {\bibfnamefont {O.~V.}\
  \bibnamefont {Gritsenko}}, \ and\ \bibinfo {author} {\bibfnamefont {E.~J.}\
  \bibnamefont {Baerends}},\ }\bibfield  {title} {\enquote {\bibinfo {title}
  {One - determinantal pure state versus ensemble kohn-sham solutions in the
  case of strong electron correlation: Ch2 and c2},}\ }\href {\doibase
  10.1007/s002140050343} {\bibfield  {journal} {\bibinfo  {journal}
  {Theoretical Chemistry Accounts}\ }\textbf {\bibinfo {volume} {99}},\
  \bibinfo {pages} {329--343} (\bibinfo {year} {1998})}\BibitemShut {NoStop}%
\end{thebibliography}%

\end{document}